\documentclass[prb,twocolumn,aps,superscriptaddress,showpacs,floatfix]{revtex4}
\usepackage{amsmath}
\usepackage{graphicx}
\begin{document}
\title{Generation of tunable Terahertz out-of-plane radiation using Josephson
vortices in modulated layered superconductors}
\author{Sergey Savel'ev}
\affiliation{Frontier Research System, The Institute of Physical
and Chemical Research (RIKEN), Wako-shi, Saitama, 351-0198, Japan}
\author{Valery Yampol'skii}
\affiliation{Frontier Research System, The Institute of Physical
and Chemical Research (RIKEN), Wako-shi, Saitama, 351-0198, Japan}
\affiliation{A. Ya. Usikov Institute for Radiophysics and
Electronics NASU, 61085 Kharkov, Ukraine}
\author{Alexander Rakhmanov}
\affiliation{Frontier Research System, The Institute of Physical
and Chemical Research (RIKEN), Wako-shi, Saitama, 351-0198, Japan}
\affiliation{Institute for Theoretical and Applied Electrodynamics
RAS, 125412 Moscow, Russia}
\author{Franco Nori}
\affiliation{Frontier Research System, The Institute of Physical
and Chemical Research (RIKEN), Wako-shi, Saitama, 351-0198, Japan}
\affiliation{Center for Theoretical Physics, Center for the Study
of Complex Systems, Department of Physics, University of Michigan,
Ann Arbor, MI 48109-1040, USA}
\date{\today}
\begin{abstract}
We show that a moving Josephson vortex in spatially modulated
layered superconductors generates out-of-plane THz radiation.
Remarkably, the magnetic and in-plane electric fields radiated are
of the same order, which is very unusual for any good-conducting
medium. Therefore, the out-of-plane radiation can be emitted to
the vacuum without the standard impedance mismatch problem. Thus,
the proposed design can be more efficient for tunable THz emitters
than previous proposals, for radiation only propagating along the
${ab}$-plane.

\end{abstract}
\pacs{74.72.Hs, 
74.78.Fk, 
41.60.-m
} \maketitle
\section{Introduction}
The recent growing interest in terahertz (THz) science and
technology is due to its many important applications in physics,
astronomy, chemistry, biology, and medicine, including THz
imaging, spectroscopy, tomography, medical diagnosis, health
monitoring, environmental control, as well as chemical and
biological identification~\cite{tera-appl}. This range of the
electromagnetic spectrum sits between 0.3 and 30 THz, which
corresponds to 1000-10 $\mu$m (wavelength), 1.25-125\,meV (energy)
or 14-1400 K (temperature). The THz gap, that is still hardly
reachable for both electronic and optical devices, covers
temperatures of biological processes and a substantial fraction of
the luminosity remanent from the Big Bang~\cite{tera-appl}.

High-temperature $\rm Bi_2Sr_2CaCu_2O_{8+\delta}$ superconductors
have a layered structure that allows the propagation of
electromagnetic waves (called Josephson plasma
oscillations~\cite{plasma,plasma-theor,plasma-exp,plasma-exp1,plasma-exp2})
with Josephson plasma frequency $\omega_J$. This is drastically
different from the strong damping of electromagnetic waves in low
temperature superconductors. The Josephson plasma frequency lies
in the THz range (see, e.g.,~\cite{bis-tera,tera2,tachiki}).
Indeed, tunable filters of THz radiation have been proposed using
the Josephson vortex lattice as a tunable photonic crystal
\cite{photonic}.
Moreover, detectors of THz radiation have been very recently
proposed using surface Josephson plasma waves \cite{surface}.
A possible way to generate THz radiation in $\rm
Bi_2Sr_2CaCu_2O_{8+\delta}$ and related compounds is to apply an
in-plane magnetic field $H_{ab}$ and an external current
$J_{\parallel c}$ perpendicular to the superconducting layers
(i.e., along the $c$-axis). Josephson vortices (JVs) induced by
$H_{ab}$ and driven fast by the {\textbf c}-axis current emit THz
radiation (e.g.,~\cite{bis-tera,tachiki}). However, it was
shown~\cite{mints-kras,cher1,cher2} that the radiation propagates
{\it only} along the plane of motion of the JVs and decays in the
$c$-direction. This THz radiation is characterized by a huge
impedance mismatch resulting in a very small fraction of THz wave
intensity emitted from the sample~\cite{tachiki}. This impedance
mismatch is a very important problem restricting possible
applications~\cite{ustin1}.

To avoid this problem, we propose a new class of THz emitters
based on JVs moving through in-plane modulated layered
superconductors, including both the strongly anisotropic
high-$T_c$ $\rm Bi_2Sr_2CaCu_2O_{8+\delta}$ single crystals and
artificial stacks of Josephson junctions (SJJ), e.g.,
Nb-Al-AlO$_x$-Nb. In-plane spatial variations of the Josephson
maximum {\textbf c}-axis current $J_c$ can be obtained by using
either irradiation of a standard $\rm Bi_2Sr_2CaCu_2O_{8+\delta}$
sample covered by a modulated mask (see, e.g.,~\cite{kwok}) or
pancake vortices controlled by an out-of-plane magnetic
field~\cite{koshelevprl}.

In order to pass through the superconductor-vacuum interface
without a significant decrease of the amplitude, the electric and
magnetic components of the propagating wave have to be of the same
order of magnitude. This feature is inherent for the out-of-plane
Josephson plasma waves (JPW), propagating both along and
perpendicular to the layers with short wavelength along the
$c$-axis. For such waves, the transmission coefficient is about
unity \cite{helm}. The out-of-plane JPW can be emitted, for
instance, by a fast moving Josephson vortex if its velocity $V$
exceeds a certain threshold value $V_{\min}$. However, this
out-of-plane Cherenkov-type radiation always completely reflects
from the sample boundary and thus cannot be emitted into the
vacuum. Indeed, the longitudinal wave vector $q$ for the Cherenkov
radiation is related to the wave frequency $\omega$ by
$q=\omega/V$ and is much larger than the maximum possible value
$\omega/c$ for waves in vacuum. This problem can be solved if the
out-of-plane Cherenkov radiation propagates through a modulated
layered superconductor. The out-of-plane Cherenkov wave
interacting with periodic inhomogeneities generates new modes with
wave vectors $q_m=q-2\pi m/a$, where $a$ is the spatial period of
the modulations and $m$ is an integer. Thus, the wave vector
$q_1=q-2\pi/a$ can meet the condition $q_1<\omega/c$ for vacuum
waves and is emitted from a sample without an impedance mismatch.
It is important to stress that the Cherenkov radiation generated
by any relativistic particle in any medium undergoes a complete
internal reflection (since $q>\omega/c$) and, thus, we propose
this general way of emitting any Cherenkov-type radiation to
vacuum. Here, we predict this out-of-plane Cherenkov radiation,
and derive the modes propagating in a modulated superconductor and
emitted into the vacuum.

Our proposal mainly concerns bulk layered superconductors, where
it is important to have radiation propagating not only along the
$ab$-planes, but also along the $c$-axis. For the case of thin
films having a thickness much smaller than the in-plane London
penetration depth (about 200 nm), the radiation damped along the
$c$-axis can also give some contribution to the waves emitted from
the wide sample side, which is parallel to the $ab$-plane. This
can even increase the fraction of the out-of-plane radiation
discussed in this paper. In other words, we show that there are
two contributions to the out-of-plane radiation along the
$c$-axis: damped waves and propagating waves. The latter one is
the real out-of-plane radiation. The former one (damped waves) can
contribute to the out-of-plane radiation only if the sample is
sufficiently thin. However, we will not consider this case in
detail because the power of the emitted radiation decreases for
decreasing thickness.

\subsection{Brief summary of the results}

For electromagnetic waves in any conducting media, the electric
field $E$ is very weak with respect to the magnetic field $H$:
$E\ll H$. Also, for in-plane radiation: $E\ll H$. Thus, only a
small fraction ($\sim\; E/H$) of the radiation can leave the
sample. This is the so-called ``impedance mismatch'' problem that
has severely limited progress in this field for years. Now, we are
also considering $c$-axis short-wavelength out-of-plane radiation.
This radiation has a strong enough in-plane electric field
$E_{\parallel}$ to overcome the superconducting--vacuum interface.
Indeed, $E_{\parallel}$ and the magnetic field both are of the
same order of magnitude, similar to the one for waves propagating
in the vacuum. This solves the impedance mismatch problem. Thus,
we propose to use $c$-axis short-wavelength out-of-plane
radiation, propagating in a {\it periodically modulated}
Bi$_2$Sr$_2$CaCu$_2$O$_{8+\delta}$ sample, to {\it overcome} the
severe {\it impedance mismatch problem} which limits the
application of layered superconductors for THz emitters.

The spatial modulations of the maximum Josephson current allows
the emitted waves to shift their wavenumbers (within a wide range
$\omega_J/V$) towards the narrow spectral window
$\sim\;\omega_J/c$ for waves propagating in the vacuum. Thus, the
emitted waves can pass the superconductor-vacuum interface within
a narrow frequency window $\sim \omega_J V/c$. This offers the
possibility to select a narrow frequency window from the
initially-broad THz radiation produced by the JVs. Also, this
should allow to achieve superradiance via the stabilization of the
square Josephson lattice moving in a periodically modulated $\rm
Bi_2Sr_2CaCu_2O_{8+\delta}$ samples\cite{unpub}. Other recent
novel ways to control vortex motion are also attracting
considerable attention \cite{vor1,vor2}.

\section{Model}
We consider an infinite layered superconductor as
in Fig.~1b. Following \cite{art}, we assume that the
superconducting layers are extremely thin, so that the spatial
variations of the phase of the superconducting order parameter and
the electromagnetic field inside the layers in the direction
perpendicular to the layers can be neglected. We choose the
$xy$-plane to be parallel to the crystallographic $ab$-plane and
the $c$-axis along the $z$-axis. Superconducting layers are
numbered by the subscript $l$. The electric $\vec E$ and magnetic
$\vec H$ fields have components, ${\vec E} = \{E_x, 0, E_z\}, \,
{\vec H} = \{0, H, 0\}$.

The gauge-invariant phase difference $\varphi_{l}$ between
$(l+1)$th and $l$th superconducting layers is described by coupled
sine-Gordon equations~\cite{nsg,art,helm},
\[
\left(1-\frac{\lambda_{ab}^2}{D^2}\;\Delta_l\right)\left(\frac{\partial^2
\varphi_{l}}{\partial t^2} + \omega_J^2(1+\mu(x))
\sin(\varphi_{l})\right)
\]
\begin{equation}\label{e12}
- \frac{c^2}{\varepsilon}\frac{\partial^2 \varphi_{l}}{\partial
x^2}=0, \qquad \mu \ll 1.
\end{equation}
Here $D$ is the spatial period of the layered structure,
$\lambda_{ab}$ is the London penetration depth along the $c$-axis,
the operator $\Delta_l$ is defined as
$$\Delta_l f_l =
f_{l+1}+f_{l-1}-2f_l,
$$
and
\begin{equation}
\omega_J = \sqrt{\frac{8\pi e D J_c}{\hbar\varepsilon}}
\end{equation}
is the Josephson frequency, $\varepsilon$ is the interlayer
dielectric constant, and the modulation factor $\mu(x)=\mu(x+a)$
is a periodic function with spatial period $a$. For simplicity, we
neglect the relaxation term related to the quasiparticle current,
which is very small at low temperatures.

The coupled sine-Gordon equations (\ref{e12}) describe both the
Josephson vortices in layered superconductors and the emitted
Josephson plasma waves. In the last case Eq.~(\ref{e12}) should be
linearized, i.e., $\sin(\varphi_{l})$ replaced by $\varphi_{l}$.

\section{Nonlocal sine-Gorgon equation}
In zero approximation with respect to the modulation $\mu$, the
travelling wave solution of Eq.~(\ref{e12}),
$\varphi_l=\phi_l(\zeta=x-Vt)$, represents the JV moving with a
constant velocity. The main phase difference $\phi=\phi_0(\zeta)$
occurs at the central junction, where nonlinearity plays a crucial
role, while equations for junctions with $l\ne 0$ can be
linearized. However, the magnetic flux of a JV spreads over a
large number of these junctions, $l \sim \lambda_{ab}/D \gg 1$.
Thus, the equation for the central junction cannot be decoupled
from others. This determines a complicated non-local structure of
the JV in layered superconductors.

Following the approach by Gurevich~\cite{gur}, the closed-form
non-local sine-Gordon equation for $\phi_0(\zeta)$ can be derived
(see Appendix) ,
\begin{equation}\label{17}
\frac{V^2}{\omega_J^{2}}\;\frac{\partial^2\phi}{\partial\zeta^2}+\sin\phi
=\frac{\lambda_J^{2}}{\pi\lambda_c}\int_{-\infty}^{\infty}
\mathrm{d}\zeta'
\,K_0\,\left(\frac{|\zeta'-\zeta|}{\lambda_c}\right)\frac{\partial^2\phi}{\partial\zeta'^2},
\end{equation}
where $K_0(x)$ is the modified Bessel function,
$$\lambda_J^2 =
\frac{c\Phi_0}{16\pi^2\lambda_{ab}J_c}$$ is the Josephson length,
$\lambda_c = c/\omega_J \sqrt{\varepsilon}$ is the penetration
depth of the magnetic field along layers, and $\Phi_0$ is the flux
quantum. Strictly speaking, equation~(\ref{17}) is derived  for a
JV moving in a weaker junction with a critical current
$J_c^{w}<J_c$ but describes qualitatively even the stack of
identical junctions (see Appendix). Equation~(\ref{17}) is reduced
to the usual local sine-Gordon equation only for $\lambda_c \ll
\lambda_J$, i.e., if the kernel $K_0$ in the integral in
Eq.~(\ref{17}) is a sharper function of $\zeta'$ than
$\partial^2\phi/\partial\zeta'^2$. For $\rm
Bi_2Sr_2CaCu_2O_{8+\delta}$ as well as for artificial layered
superconductors, the opposite strongly non-local limit,
$\lambda_c\, \gg \, \lambda_J$, is realized. In this case, a
soliton-like solution,
\begin{equation}
\phi=\pi+2\arctan\left(\frac{2x}{L}\right),
\end{equation}
of Eq.~(\ref{17}) was obtained in Ref.~\cite{gur} for a fixed JV,
$V=0$, with the soliton size
\begin{equation}
L\;=\;\frac{2\lambda_J^2}{\lambda_c}\;=\;\frac{\lambda_c
D}{\lambda_{ab}}\;\ll\;\lambda_J.
\end{equation}
It is important to stress that the soliton size $L$ (rather than
$\lambda_J$) coincides with the well-known estimate of the JV core
($L=\gamma D$, $\gamma=\lambda_c/\lambda_{ab}$) in a layered
superconductor. Analytical and numerical analysis proves that the
size $L(V)$ of the moving JV remains of the same order at any
allowed vortex velocities
\begin{equation}
V\;<\;V_c\ \sim\ \omega_J
L\;=\;\frac{cD}{\lambda_{ab}\sqrt{\varepsilon}}.
\end{equation}
 Note that, due to
non-locality, the maximum vortex velocity $V_c$ in a layered
superconductor is much smaller than the maximum vortex velocity
$c_{\rm sw}=\lambda_J\omega_J$ (so-called Swihart velocity) in a
single junction \cite{barone}.

\section{Josephson plasma waves}
In homogeneous ($\mu=0$) layered
superconductors, the linearized coupled sine-Gordon equations
admit wave solutions of the form,
\begin{equation}\label{e14}
\varphi_{l} = \varphi_0\exp\{i[qx - \omega t + k(q,\omega)lD]\},
\end{equation}
with the dispersion relation,
\begin{equation}\label{e15}
\sin^2\left(\frac{kD}{2}\right)=\frac{D^2}{4\lambda_{ab}^2}\left(\frac{c^2q^2}{\varepsilon
(\omega^2-\omega_J^2)} -1 \right),
\end{equation}
for the transverse wave vector $k(q,\omega)$. This relation
coincides with the spectrum obtained in~\cite{helm} in the
particular limit where the breaking of the charge neutrality
effect (that we can easily take into account) is neglected.

The electric and magnetic fields exhibit the same spatio-temporal
dependence as in Eq.~(\ref{e14}) for the phase difference
$\varphi_l$; while their amplitudes, $H_0$, $E_{x0}$, and
$E_{z0}$, are related to $\varphi_{0}$ via
\begin{equation}
\left[1+\frac{4\lambda_{ab}^2}{D^2}\sin^2\left(
\frac{kD}{2}\right) \right] H_0(q,\omega)=\frac{iq\Phi_0}{2\pi D}
\varphi_0(q,\omega),
\end{equation}
\begin{equation}
E_{z0}(q,\omega)=-\;\frac{i\omega\Phi_0}{2\pi c D}\;
\varphi_0(q,\omega),
\end{equation}
\begin{equation}\label{ex}
E_{x0}(q,\omega)=\frac{i\omega \lambda_{ab}^2}{c D} [1-\exp(-
ikD)]H_0(q,\omega).
\end{equation}

According to Eq.~(\ref{e15}), the JPW can propagate if $\omega
> \omega_J$. For homogeneous layered superconductor, JVs moving
with constant velocity $V$ can excite waves with $q=\omega/V$
(Cherenkov radiation~\cite{bis-tera,mints-kras,cher1,cher2}).
Substituting $\omega=qV$ in Eq.~(\ref{e15}) and noticing that the
right-hand side of this equation is less than unity, we obtain the
limiting vortex velocity
\begin{equation}
V_{\min}=\frac{cD}{2\lambda_{ab}\sqrt{\varepsilon}}.
\end{equation}
If the vortex velocity is larger than this threshold value,
$V>V_{\min}$, then  $\mathrm{Im}(k)=0$ and the {\it
out-of-plane\/} waves can be excited. The characteristic angle
$\theta$ of the propagating radiation (Cherenkov cone)
\begin{equation}
\tan\theta=\left(\frac{\pi\sqrt{V^2-V_{\min}^2}}{D\omega_J}\right)
\end{equation}
is determined by three standard conditions: (i) Eq.~(\ref{e15}),
(ii) $\omega=qV$, and (iii) the minimum wavelength $k^{-1}\lesssim
D/\pi$. For $V=V_{\min}$ the Cherenkov cone is obviously closed,
$\theta=0$. At lower vortex velocities, $V<V_{\min}$, the
Cherenkov radiation with $\mathrm{ Im}(k)\ne 0$ (decaying from the
central junction)~\cite{mints-kras} can propagate only along the
{\textrm ab}-plane~\cite{trans} (see also \cite{malom} where the
Cherenkov radiation in two coupled junctions is studied).

As was shown above, the maximum vortex velocity $V_c\sim \omega_J
L$, at which the moving soliton is stable, is of the same order as
$V_{\min}$. Below, the out-of-plane Cherenkov radiation generated
by a vortex moving in a junction weaker than others is derived
(while the question if such radiation exists in a system of
identical junctions remains open ~\cite{bis-tera}). A subset of
weaker intrinsic Josephson junctions in $\rm
Bi_2Sr_2CaCu_2O_{8+\delta}$-based samples can be made using either
(i) the controllable intercalation technique \cite{two-dif-layer},
(ii) Chemical Vapor Deposition (CVD) (see, e.g., \cite{CVD}), or
(iii) via the admixture of $\rm Bi_2Sr_2Cu_2O_{6+\delta}$ and $\rm
Bi_2Sr_2Ca_2Cu_3O_{10+\delta}$ \cite{admixture}. Also, such a
system can be easily created using artificial stacks of layers of
low temperature superconductors.

\section{Out-of-plane Cherenkov radiation}
In order to find the relation between the amplitudes of the
emitted waves inside a superconductor and the phase difference,
$\phi$, in the central junction, we use the standard equation
\cite{gur,barone}:
\begin{equation}\label{10}
\frac{\mathrm{d}\phi}{\mathrm{d}\zeta} = \frac{8\pi ^2
\lambda_{ab}^2}{c \Phi_0}\{J_x^+ (\zeta) - J_x^- (\zeta)\},
\end{equation}
where $J_x^{\pm}$ are the currents at the top and bottom edges of
the central contact. Equation (\ref{10}) is valid if $D\ll
2\lambda_{ab}$, that is, if the magnetic flux through the central
junction is small compared to $\Phi_0$. In other words, the
gradient of the gauge invariant phase difference $\phi$ along the
central junction occurs due to the gradient of the phase
$\chi^{\pm}$ of the superconducting order parameter in the layers
forming the central junction rather than the trapped magnetic
flux. Using Eq. (\ref{10}) and the Maxwell equation $\partial
H/\partial z= -4\pi J_x/c $, we obtain
\begin{equation}\label{12}
H_0(q,\omega=qV)=\ -\ \frac{i \Phi_0 D q}{4\pi\lambda_{ab}^2
(1-\exp(-ik(q,qV)D))}\phi(q)\ ,
\end{equation}
where $\phi(q)$ is the Fourier transform of $\phi(\zeta)$.

When $\phi(q)$ is known, Eq.~(\ref{12}) determines the magnetic
field distribution of the emitted out-of-plane Cherenkov
radiation. For simplicity, we use the Fourier transform
$\phi(q)=\phi_0(q)$ of the solution for a fixed vortex,
\begin{equation}
\phi(q)=-2\pi \frac{i\exp(-|q|L)}{q}\ .
\end{equation}
 Using this last equation and
integrating $H_0(q,qV)\exp{(iqx-iqVt+ik(q,qV)z)}$ over
$q_{\min}<q$ for the travelling out-of-plane waves we derive the
expression for the magnetic field $H_{\rm Cher}$ of the radiation:
\[
H_{\rm Cher}(x,z,t)=\frac{\Phi_0 D}{2\pi\lambda_{ab}^2}
\int_{q_{min}}^\infty\frac{\mathrm{d}q}{1-\exp[-{\mathrm i}k(q)D]}
\]
\begin{equation}\label{37}
\times\exp(-qL)\sin[q(x-Vt)+k(q,qV)|z|],
\end{equation}
where
\begin{equation}
q_{\min}\;=\;\frac{\omega_J}{\sqrt{V^2-V_{\min}^2}}\ .
\end{equation}
 The magnetic field
distribution (\ref{37}) in the emitted Cherenkov waves is shown in
Fig. 1a.

Due to the rather unusual dispersion relation (\ref{e15}), i.e.,
the decrease of $k(q,\omega=qV)$ with increasing $q$, as well as
due to the spatial extension of a vortex, the generated
electromagnetic waves are located {\it outside} the Cherenkov cone
(Fig.~1a), which is drastically different from the Cherenkov
radiation of a fast (point-like) relativistic particle. The new
type of radiation predicted here could be called {\it
outside-the-cone} Cherenkov radiation.

\begin{figure}[!htp]
\begin{center}
\begin{center}
\vspace*{-0.2cm}
\includegraphics*[width=8.7cm]{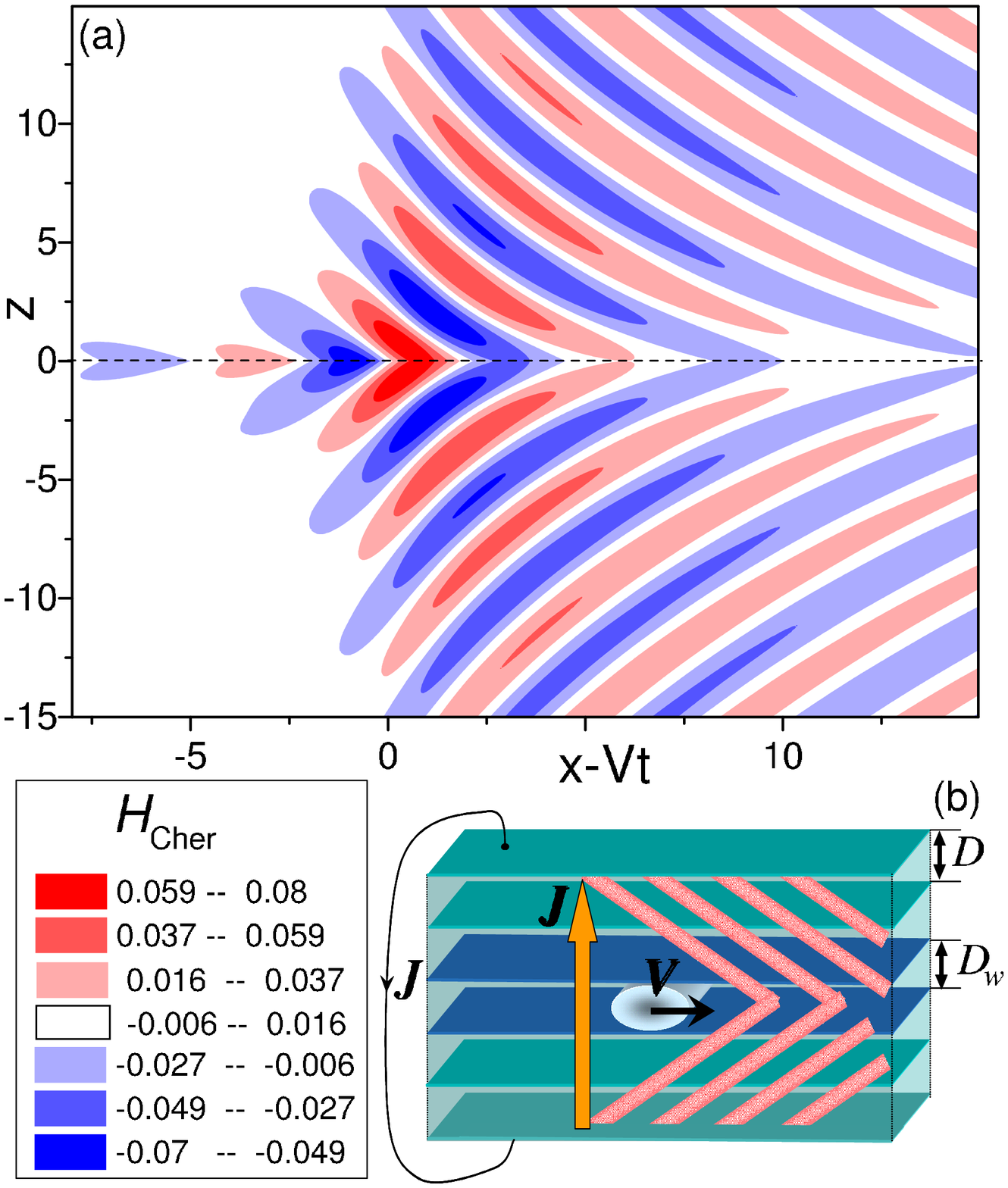}
\end{center}
\vspace*{-0.5cm} \caption{Cherenkov radiation generated by a fast
Josephson vortex (located at $x=Vt$) moving in a weaker junction
with the critical current $J_c^w<J_c$ and the junction thickness
$D_w$. (a) Magnetic field distribution $H(x-Vt,z)$ in units of
$\Phi_0/2\pi\lambda_c\lambda_{ab}$ ($\sim H_{\rm c1}^J\;=$ the
first critical magnetic field for the Josephson vortices) for
$J_c^w/J_c=0.2$, $D_w/D=1.2$, $V/V_{\max}=0.9$ is calculated using
(\ref{37}). The ``running'' coordinate, $x-Vt$, is measured in
units of $\gamma D / (\pi\sqrt{v^2\beta^2-1})$, while the
out-of-plane coordinate $y$ is normalized by
$D/(\pi\sqrt{v^2\beta^2-1})$, where $\beta= J_cD_w/J_c^wD$. The
moving vortex emits radiation propagating forward. This radiation
forms a cone determined by the vortex velocity $V$. (b) Geometry
of the problem: in a weaker junction (located between the two blue
superconducting planes in (b)) a ${c}$-axis current $J_{\parallel
c}$ drives a Josephson vortex with velocity $V$, which is higher
than the minimum velocity $V_{\min}$ of the propagating
electromagnetic waves. Red strips in (b) schematically show
outside-the-cone Cherenkov radiation.}
\end{center}
\end{figure}

\section{Impedance mismatch} It is important to emphasize that
a fast moving vortex emits mostly the JPWs with short out-of-plane
lengths, i.e., with $k(q)$ about $k(q_{\min})=\pi/D$ (see
Fig.~1a). According to Eq.~(\ref{ex}), the ratio $E_x/H$ is about
\begin{equation}
\beta=\frac{2\omega_J\lambda_{ab}^2}{cD}\sim 1.1
\end{equation}
 for these short waves.
For this estimate we use commonly accepted parameters for Bi2212
samples ($\omega_J/2\pi\sim$ 1 THz, depending on doping and
temperature, $\lambda_{ab}=2000$\AA\,, and $D=15$\AA). This
result, unusual for conducting media, suggests that there is no
impedance mismatch for the out-of-plane radiation.

However, as was mentioned above, the Cherenkov radiation can never
pass through the sample boundary because it has a large
longitudinal wave vector $q=\omega/V\gg \omega/c$. In order to
decrease the longitudinal wave vector $q$, the spatial modulations
of the critical current can be used. To analyze this analytically
we use the perturbation scheme $H=H^{(0)}+\mu H^{(1)}+..$ with
$\mu(x)=\mu\cos(2\pi x/a)$, $\mu\ll 1$, in Eq.~(\ref{e12}).

In zeroth-order approximation, at the sample boundary, the mode
\begin{equation}
H_0^{+}(x,z,t)=H_{\rm 0,0}^{+}\cdot\exp\left[iqx-iqVt+
ik\left(q,qV\right)z\right],
\end{equation}
generated by a vortex completely reflects back to the
superconductor sample as
\begin{equation}
H_0^{-}(x,z,t)=H_{\rm 0,0}^{-}\cdot\exp\left[iqx-iqVt-
ik\left(q,qV\right)z\right],
\end{equation}
generating the decaying wave in vacuum
\begin{equation}
H_{\rm damp}(x,z,t)=H_{\rm vac}^{\rm
damp}\cdot\exp[iqx-iqVt-\kappa_v z]
\end{equation}
with a damping coefficient
\begin{equation}
\kappa_v=q\sqrt{1-\frac{V^2}{c^2}}.
\end{equation}
The amplitudes $H_{\rm 0,0}^{+}$, $H_{\rm 0,0}^{-}$, and $H_{\rm
vac}^{\rm damp}$ of the modes are determined by the continuity of
the electric and magnetic fields at the sample boundary.

To the first approximation, using the linear relations between
$\phi_l$ and $H_l$, as well as equation (\ref{e12}), we obtain
the following equation:
\[
\left(1-\frac{\lambda_{ab}^2}{D^2}\Delta_l\right)
\left(\frac{\partial^2 H_{l}^{(1)}}{\partial t^2} + \omega_{J}^{2}
H_{l}^{(1)} \right)
\]
\begin{equation}\label{first}
- \frac{c^2}{\varepsilon}\frac{\partial^2 H_{l}^{(1)}}{\partial
x^2}=-\mu\cos\left(\frac{2\pi
x}{a}\right)\left(1-\frac{\lambda_{ab}^2}{D^2}\;\Delta_l\right)\omega_J^2
\left(H_0^{+}+H_0^{-}\right).
\end{equation}
The solution of this equation consists of the sum of the following
modes:
\[
H_{n,m}^{\pm}(x,z,t)=
\]
\begin{equation}
H_{n,m}^{\pm}\cdot\exp\left[i\left(q-\frac{2\pi
n}{a}\right)x-iqVt\pm ik\left(q-\frac{2\pi
m}{a},qV\right)z\right],
\end{equation}
where $H_{\rm 1,0}^{\pm}$ and $H_{\rm -1,0}^{\pm}$ are forced
waves and $H_{\rm 1,1}^{\pm}$ and $H_{\rm -1,-1}^{\pm}$ are
eigensolutions. The modes $H_{\rm -1,0}^{\pm}$ are also reflected
from the boundary, while the modes $H_{\rm 1,0}^{\pm}$ generate
the wave
\begin{equation}
H_{\rm vac}^{\rm
prop}\cdot\exp\left[i\left(q-\frac{2\pi}{a}\right)x-iqVt+
ik_vz\right],
\end{equation}
propagating in vacuum with wave vector
\begin{equation}
k_v=\sqrt{\frac{q^2V^2}{c^2}-\left(q-\frac{2\pi}{a}\right)^2}
\end{equation}
The continuity of $E_x$ and $H$ for the longitudinal wave vector
$q-2\pi/a$ define the amplitudes of the waves $H_{\rm 1,0}^{\pm}$,
$H_{\rm 1,1}^{-}$, and $H_{\rm vac}^{\rm prop}$, as well as the
transition coefficient $T$ for the emitted waves from the sample,
\begin{equation}
T\ =\ \frac{E_{x\, \rm vac}^{\rm prop}}{H_{0,0}^{+}}\ =\ -\ \mu\;
\beta\;\frac{v(v^2-1)p^2\sqrt{p^2-1}}{[1+v^2(p^2-1)]^2}
\end{equation}
for $|p-2\pi/aq_{\min}| < V/c$ and $T=0$ otherwise. Here,
$p=q/q_{\min}$ and $v=V/V_{\min}$ are the dimensionless
longitudinal wave vector and vortex velocity.

Because of $\beta\sim 1$, the transmission coefficient $T\sim 1$
for out-of-plane radiation in a narrow frequency region
\begin{equation}
\Delta\omega \; \sim \; \frac{ \omega_JV }{ c } \; \ll \;
\omega_J.
\end{equation}
The quite narrow window of the transmitted waves occurs due to the
broad spectrum,
\begin{equation}
\Delta\omega_{\rm Cherenkov}\ \sim\ \omega_J,\ \ \ \ \ \Delta
q_{\rm Cherenkov}\ \sim\ \Delta\omega/V,
\end{equation}
of the Cherenkov radiation emitted by a Josephson vortex with
respect to the spectrum of waves propagating in the vacuum:
\begin{equation}
\Delta q_{\rm vac}\ \sim\ \omega_J/c\ \ll\ \Delta q_{\rm
Cherenkov}\ .
\end{equation}
 A periodic
spatial modulation can shift all the wave vectors towards the
spectral window of the vacuum waves, while it cannot affect the
width of the spectrum. Thus, the surface cuts a narrow strip from
the broad Cherenkov radiation, allowing these waves to pass the
interface. Note that changing the period $a$ (e.g., changing the
distance between pancake vortices via the $c$-axis magnetic field)
allows to tune the frequency window for the emitted radiation. In
order to make the frequency window $\Delta\omega$ wider, one can
employ, e.g., appropriate aperiodic modulations
$\mu(x)=\int_{\kappa_1}^{\kappa_2}\cos\kappa x\, \mathrm{d}\kappa$
with $\kappa_1$ and $\kappa_2$ incommensurate. Also, $\mu(x)$
could be modulated as a 1D quasicrystal\cite{misko}.

\section{Conclusions}
We propose how to generate out-of-plane THz radiation in a
controllable frequency range. We show that the standard severe
mismatch problem can be overcome here for out-of-plane radiation
using spatially modulated samples. Moreover, recent studies
\cite{super0,super} of Josephson vortex arrangements in small
samples (there, the interaction of vortices with sample boundaries
acts similar to an additional potential) suggest a way to obtain a
square vortex lattice, which is important for superradiance
\cite{lobb}. Thus, spatial modulations of $J_c$ can result in a
more ordered vortex flow or even a flowing square vortex lattice
generating superradiance. Of course this problem requires more
detailed studies, which will be presented in the future.

\section{Acknowledgements}
We gratefully acknowledge conversations with M. Gaifullin, A.
Koshelev, M. Tachiki, A.V. Ustinov, and partial support from the
NSA and ARDA under AFOSR contract No. F49620-02-1-0334, and by the
NSF grant No. EIA-0130383.

\section*{Appendix: Derivation of the nonlocal equation for a Josephson vortex}

We consider the distributions of the gauge-invariant phase
differences and the magnetic field of a single vortex located in
the central junction. We assume that the main phase difference is
across this junction, whereas $\phi_l$ are small across other
junctions. Such an approximation is, strictly speaking, correct
for a weaker contact with $J_c^{w}<J_c$. This approach is also
applicable for qualitative analysis for the set of identical
junctions. In particular, it provides correct asymptotic behavior
for $\phi_l$ when $l\gg 1$. Thus, we can use the linearized Eqs.
(\ref{e12}) for all junctions except the junction where a JV is
localized. It is important to stress that the magnetic field and
the corresponding phase difference distributions generated by the
vortex are extended over many junctions in the layered medium.
Therefore, we can use Eqs. (\ref{e12}) in the continuum form,
which reads
\begin{equation}\label{1}
\left( 1-\lambda_{ab}^2\frac{\partial ^2}{\partial z^2}\right)
\left( \omega_J^2 + V^2 \frac{\partial ^2}{\partial
\zeta^2}\right)\phi (\zeta,z) -
\frac{c^2}{\varepsilon}\frac{\partial ^2 \phi}{\partial \zeta^2}
=0.
\end{equation}

Using the Fourier transform
\begin{equation}\label{4}
\phi (q,z) = \int_{-\infty}^{\infty} \mathrm{d}\zeta \exp(-\mathrm
i q \zeta)\phi (\zeta,z)\;,
\end{equation}
we obtain the solution of Eq.~(\ref{1})
\begin{equation}\label{5}
\phi (q,y) = \phi(q,0) \exp[\;\mathrm{i}\,\mathrm{sign}(q)\, k(q)
|z|\;],
\end{equation}
where $\mathrm{sign}(q)=1$ if $q>0$ and -1 if $q<0$, the wave
vector $k(q)$ is defined by Eq.~(\ref{e15}) for $kD\ll 1$,
$\omega=qV$ and $V\ll c/\varepsilon$
\begin{equation}\label{6}
k(q) = \frac{1}{\lambda_{ab}} \left( \frac{\omega_J^2 + c^2 q^2
/\varepsilon}{q^2 V^2 - \omega_J^2}\right)^{1/2}.
\end{equation}

Using the Maxwell equation and considering the Josephson and
displacement currents, we derive relation between $H$ and $\phi$:
\begin{equation}\label{7}
-\; \frac{\partial H}{\partial \zeta} = \frac{4\pi}{c}\left(J_c
\phi +\frac{\hbar\varepsilon V^2}{8\pi e s}\frac{\partial ^2
\phi}{\partial \zeta ^2} \right).
\end{equation}
The Fourier transform of Eq. (\ref{7}) results in
\begin{equation}\label{8}
H(q,z) = \;-\; \frac{8\pi \mathrm{i}}{cq}\left(J_c -
\frac{\hbar\varepsilon V^2 q^2}{8\pi e s} \right) \phi(q,z).
\end{equation}
Thus, the dependence of the magnetic field $H(q,z)$ on the
transverse coordinate $z$ obeys the same law as the phase
difference $\phi(z)$,
\begin{equation}\label{9}
H(q,z) = H(q,0)\exp[\;\mathrm{i}\,\mathrm{sign}(q)\, k(q) |z|\;].
\end{equation} Next, using Eq. (\ref{10}) and the Maxwell equation
$\partial H/\partial z= - 4\pi/cJ_x$, we obtain the relation
between Fourrier components of the magnetic field and the phase
difference $\phi$:
\begin{equation}\label{ei12}
H(q,0)\ =\
-\frac{\Phi_0}{4\pi\lambda_{ab}^2}\frac{q}{k(q)}\phi(q).
\end{equation}

In order to express the $z$ component of the current in layered
media in terms of the phase difference $\phi$ in the central
junction, we use Eqs. (\ref{6}), (\ref{8}), (\ref{9}), and
(\ref{ei12}). Performing the reverse Fourier transformation, we
obtain
\[
J_z(\zeta,z)=\frac{c\Phi_0}{16\pi^2\lambda_{ab}}
\int_{-\infty}^{\infty}\frac{\mathrm{d}q}{2\pi}q^2\phi(q)
\]
\begin{equation}\label{13}
\times\sqrt{\frac{\omega_J^2-q^2V^2}{\omega_J^2+q^2c^2/\varepsilon}}
\exp(\mathrm{i}q\zeta+\mathrm{i}\,\mathrm{sign}(q)\,k(q)|z|).
\end{equation}
Below we consider the case where the main contribution to the
integral in Eq. (\ref{13}) comes from the region
\begin{equation}\label{14}
q^2\, \ll \, \frac{\omega_J^2}{V^2},
\end{equation}
which is valid if the vortex velocity is smaller than $
\lambda_J^2\omega_J/\lambda_c$.

Substituting instead of $\phi(q)$ its coordinate Fourier transform
$\phi(\zeta')$ we find the expression for the current component
$J_z$ at the edge of the central junction, $z=0$. Performing the
integration over $q$ and taking into account the inequality Eq.
(\ref{14}), one gets
\begin{equation}\label{15}
J_z(\zeta,0)=\frac{c\Phi_0}{16\pi^3\lambda_{ab}\lambda_c}\int_{-\infty}^{\infty}
\mathrm{d}\zeta'
K_0\left(\frac{|\zeta'-\zeta|}{\lambda_c}\right)\frac{\partial^2\phi}{\partial\zeta'^2},
\end{equation}
where $K_0(x)$ is the modified Bessel function of the zero order.
Equating the current Eq. (\ref{15}) to the sum of Josephson and
displacement currents in the central junction, we obtain the
non-local sine-Gordon equation for the fluxon presented in the
text. Note, that the obtained equation for $\phi$ is similar to
the non-local sine-Gordon equation obtained by Gurevich for the
fluxon in a single Josephson junction between isotropic
superconductors \cite{gur}.

\end{document}